\documentclass[aps,prb,preprint,superscriptaddress]{revtex4-1}
\usepackage{graphicx}
\usepackage{bm}
\begin{document}
\title{Topological damping Rashba spin orbit torque in ballistic magnetic domain walls}
\author{D. Wang}
\email{wangdaowei@sztu.edu.cn}
\affiliation{College of Engineering Physics, Shenzhen Technology University, Guangdong 518118, P. R. China}
\author{Yan Zhou}
\email{zhouyan@cuhk.edu.cn}
\affiliation{School of Science and Engineering, The Chinese University of Hong Kong, Shenzhen, Guangdong 518172, P. R. China}
\begin{abstract}
Rashba spin orbit torque derived from the broken inversion symmetry at ferromagnet/heavy metal interfaces has potential application in spintronic devices. In conventional description of the precessional and damping components of the Rashba spin orbit torque in magnetization textures, the decomposition coefficients are assumed to be independent of the topology of the underlying structure. Contrary to this common wisdom, for Schr\"{o}dinger electrons trespassing ballistically across a magnetic domain wall, we found that the decomposition coefficient of the damping component is determined by the topology of the domain wall. The resultant damping Rashba spin orbit torque is protected by the topology of the underlying magnetic domain wall and robust against small deviations from the ideal domain wall profile. Our identification of a topological damping Rashba spin orbit torque component in magnetic domain walls will help to understand experiments on current driven domain wall motion in ferromagnet/heavy metal systems with broken inversion symmetry and to facilitate its utilization in innovative device designs.
\end{abstract}
\date{\today}
\maketitle

One main theme in the field of nanomagnetism is to search for new approaches to realize fast and energy efficient manipulation of magnetic state, rather than using the conventional magnetic field. In the past three decades, several promising candidates, such as electric field \cite{Ohno}, laser pulses \cite{Kirilyuk} and spin current through the spin transfer torque (STT) \cite{Berger96,Slonczewski96,Li04,Thiaville05}, were proposed. A recent development along this line is the emergence of the Rashba spin orbit torque (RSOT) in magnetic systems without inversion symmetry. In a simple picture \cite{Gambardella}, the electric field along the symmetry breaking direction is equivalent to a magnetic field, dubbed the Rashba field, in the rest reference frame of an electron in motion. Due to the $s$-$d$ exchange between the local and itinerant spin degrees of freedom, the Rashba field is transformed into the RSOT acting on the local magnetization.

When it was first proposed, only the precessional component \cite{Manchon,Garate,Obata,Matos-Abiague} of the RSOT, corresponding to the torque caused by an effective Rashba field acting on the local magnetization, was considered. Upon considering the impurity and spin-flip scattering, an additional damping torque in accordance with the effective Rashba field can arise \cite{Wang12}. Subsequent theoretical investigations were devoted to exposition of the physics of the RSOT, adopting different approaches and considering sample geometries with finite extension \cite{kim13,Ortiz-Pauyac,Wang14,Pesin,van der Bijl,Freimuth13,Haney13,Wimmer16}. However, most of the previous theoretical investigations focus on the case of uniform magnetization distribution or slowly varying magnetization textures, the more important case of magnetic domain walls (DWs), which will be the focus of the current work, is almost not touched upon.

The topological description of electron transport in periodic potentials appears naturally by considering the geometric Berry phase \cite{Berry} of itinerant electrons. In the simplest case of one dimensional (1D) motion of electrons, it leads to the Zak phase \cite{Zak}, and the Thouless-Kohmoto-Nightingale-den Nijs (TKNN) invariant \cite{Thouless} for two dimensional (2D) motion. The Berry phase is generically caused by the existence of a gauge field \cite{Bruno}, which is given by the spatial variation of the periodic modulation wave function in the case of Bloch electrons. In the presence of spin orbit interaction and a background magnetic field, which is generated by a magnetic DW, the itinerant electrons will also experience a spatially varying, emergent gauge field. By analogy with the TKNN invariant and the Zak phase, we speculate that topological phase factors should arise for electrons traversing cross a DW. Actually, the effect of spatially varying magnetization on the motion of electrons was already discussed theoretically by Bruno et al. \cite{Bruno04}. Whether a similar topological effect will emerge in RSOT remains a question.

For a simple demonstration of the physics, we will use the following minimal model Hamiltonian to study the magnetization dynamics of itinerant electrons confined to the interface between a ferromagnet and a heavy metal \cite{Obata,Manchon,Garate,Matos-Abiague},
\begin{equation}
H = \frac{\textbf{p}^2}{2 m_e}  + \mu_B \bm{\sigma} \cdot \textbf{M} + \frac {\alpha_R} {\hbar} \bm{\sigma} \cdot (\textbf{p} \times \hat{z}).
\label{hamil}
\end{equation}
$\textbf{p}= -i \hbar \nabla$ is the momentum operator, $m_e$ is the electron mass, $\hbar$ is the Planck constant divided by 2$\pi$, and $\mu_B$ is the Bohr magneton. $\alpha_R$ is the Rashba constant, which measures the degree of the inversion symmetry breaking \cite{Bychkov84}. We consider only the motion of the electrons in the interface, which is a 2D $xy$ plane in our coordinate system, since previous density functional theory investigation found that the RSOT is primarily an interface effect \cite{Freimuth13}. The third term in the Hamiltonian (\ref{hamil}) is the Rashba spin orbit interaction term, showing that the main effect of the broken inversion symmetry is to introduce an effective in-plane magnetic field, which is everywhere tangential to the in-plane linear momentum $\textbf{p}$. $\bm{\sigma} = \hat{x}\sigma_x + \hat{y}\sigma_y + \hat{z} \sigma_z$ is a vector in the spinor space where $\sigma_x$, $\sigma_y$ and $\sigma_z$ are the Pauli matrices, and $\hat {x}$, $\hat {y}$ and $\hat {z}$ are unit vectors along the $x$, $y$ and $z$ directions, respectively. The Hamiltonian (\ref{hamil}) gives the energy of conduction electrons interacting through the $s$-$d$ exchange interaction with the local magnetization $\textbf{M}$. In our model treatment, we consider only the itinerant Hamiltonian as given in Eq. (\ref{hamil}), while the local magnetic moments are assumed to be static, as described by $\textbf{M}$. The variation of the vector $\textbf{M}$ inside magnetization textures is used to provide an effective 'exchange' field for the itinerant magnetization.

The Walker DW profile \cite{Schryer74} considered for the study of the RSOT is characterized by an angle $\theta$ through the expression $\textbf{M} = M (\hat{x}\sin\theta + \hat{z} \cos\theta)$ with $\cos \theta = - q \tanh (x/\lambda)$ and $\sin \theta = \chi \,\mbox{sech} \, (x/\lambda)$, where $\lambda = \sqrt{A/K}$ is the DW width. $A$ is the exchange constant and $K$ the anisotropy constant of the ferromagnet. For a general description, we consider explicitly the charge $q$ and chirality $\chi$ of a DW \cite{Braun}. Using the time dependent Pauli-Schr\"{o}dinger equation $i \hbar \partial \psi/ \partial t = H \psi$ for the spinor wave function $\psi$, the equation of motion for the spin density $\textbf{s} = \psi^ \dagger \bm{\sigma} \psi$ of conduction electrons is given by
\begin{equation}
\frac{2 m_e} {\hbar} \frac {\partial \textbf {s}} {\partial t} = \nabla \cdot \textbf {Q} +  2k_B^2 \hat {M} \times \textbf {s} + \bm {\tau},
\label{eom_mag}
\end{equation}
where the spin current density is defined as
\begin{equation}
\textbf{Q} = i (\psi ^ \dagger \nabla \bm{\sigma} \psi - \nabla \psi ^ \dagger \bm{\sigma} \psi) + k_\alpha \epsilon _{ij3} \hat {i} \hat {j} \psi ^ \dagger \psi.
\label{density_q}
\end{equation}
$\epsilon _{ijk}$ is the antisymmetric Levi-Civita symbol and a summation over repeated indices is implied in the expression for $\textbf{Q}$. A substitution of $x$, $y$ and $z$ by numbers 1, 2 and 3 is made to compactify the expression. The parameter $k_B$ is related to the Zeeman energy splitting $\hbar^2 k_B^2/2 m_e = \mu_B M$, and the constant $k_\alpha = 2 m_e \alpha_R/\hbar^2$ is an effective wave number characterizing the strength of the Rashba interaction. $\hat {M}$ is a unit direction vector for the local magnetization, $\hat {M} = \textbf {M}/M$. The precessional term $\bm {\tau}$ follows directly from the Rashba term in Eq. (\ref{hamil}), and is given by
\begin{equation}
\bm {\tau} (\textbf{k}, \bm {\rho}) = 2k_\alpha \Im(\hat {z} \psi ^\dagger \bm {\sigma} \cdot \nabla \psi - \psi ^\dagger \sigma_z \nabla \psi).
\label{tau}
\end{equation}
For later convenience, the momentum and position dependence of $\bm {\tau}$ is explicitly written out in Eq. (\ref{tau}). Our equation of motion for the spin density is identical in form to a previous result \cite{Haney10}, if the angular momentum operator is replaced by the Rashba field operator considered here. However, this connection is superficial, as the dynamics for the angular momentum are not considered here.

With Eq. (\ref{eom_mag}), it is obvious that the itinerant magnetization dynamics is governed by three torques. The first term on the right hand side of Eq. (\ref{eom_mag}) corresponds to the spin current torque acting on the itinerant magnetization, which is just the divergence of the spin current density. In the ground state, the spin current torque reduces to the exchange torque for magnetization textures, which is proportional to $\hat{m} \times \nabla^2 \hat{m}$, with $\hat {m}$ an unit vector for the itinerant magnetization. The second term describes the torque originating from the static local magnetization, whose net effect can be viewed as an effective \textit{s}-\textit{d} exchange field acting on the itinerant magnetization. The Rashba term in the Hamiltonian $H$ gives rise to the last torque on the right hand side of Eq. (\ref{eom_mag}). In equilibrium, this Rashba torque has a form identical to the Dzyaloshinskii-Moriya torque \cite{Dzyaloshinskii,Moriya,Bogdanov89,Bogdanov94,DMfield}. If a steady state electronic current is allowed to flow, the spin current torque and the Rashba torque transform into the conventional STT and RSOT, respectively. In the current carrying steady state, there is no time variation of the itinerant magnetization. Hence the various torques on the right hand side must sum to zero. Due to this torque balance, the torque induced by the spin accumulation, which corresponds to the second term on the right hand side of Eq. (\ref{eom_mag}), contains both the STT and RSOT contributions.

Eq. (\ref{tau}) gives only the RSOT for a single Bloch state in the momentum space. Using the relaxation time approximation \cite{Ashcroft}, the physical RSOT induced in the presence of an electric field $E$ along the $x$ direction can be obtained through an integration in the momentum space as
\begin{equation}
\bm {\tau} (\bm {\rho}) = - \frac{e E \tau_0} {(2\pi)^2\hbar} \oint d\varphi\, k_x\, \bm {\tau} (\textbf{k}, \bm {\rho}),
\label{atau}
\end{equation}
where $\tau_0$ is the relaxation time constant, $e$ the electron charge, and $\varphi$ the angle of the wave vector relative to the $x$-axis. As the temperature is assumed to be absolute zero, the integration is confined to the 2D Fermi surface, which is a circle.
\begin{figure}\centering
\begin{minipage}[c]{0.45\linewidth}
\includegraphics[width=\linewidth]{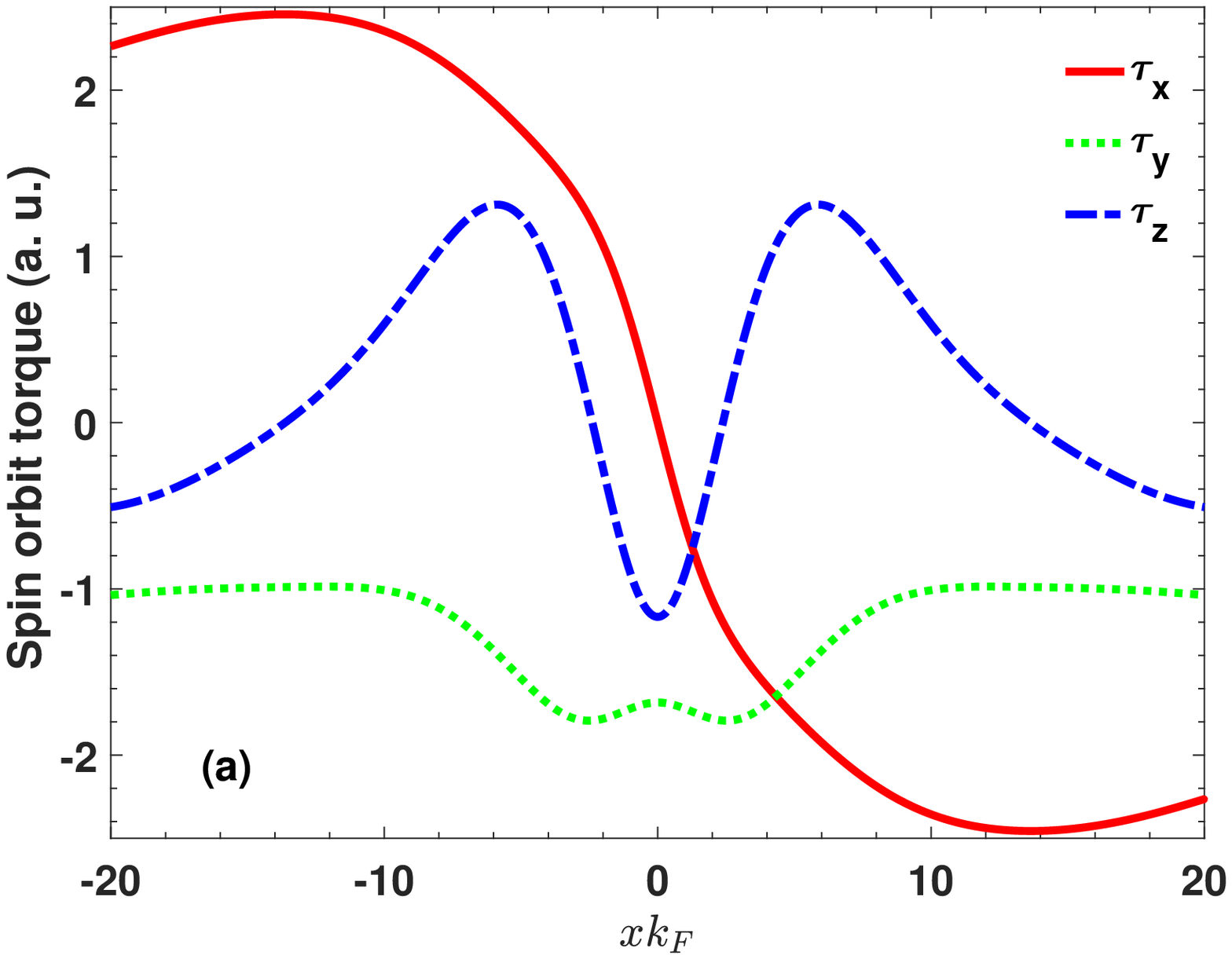}
\end{minipage}
\begin{minipage}[c]{0.45\linewidth}
\includegraphics[width=\linewidth]{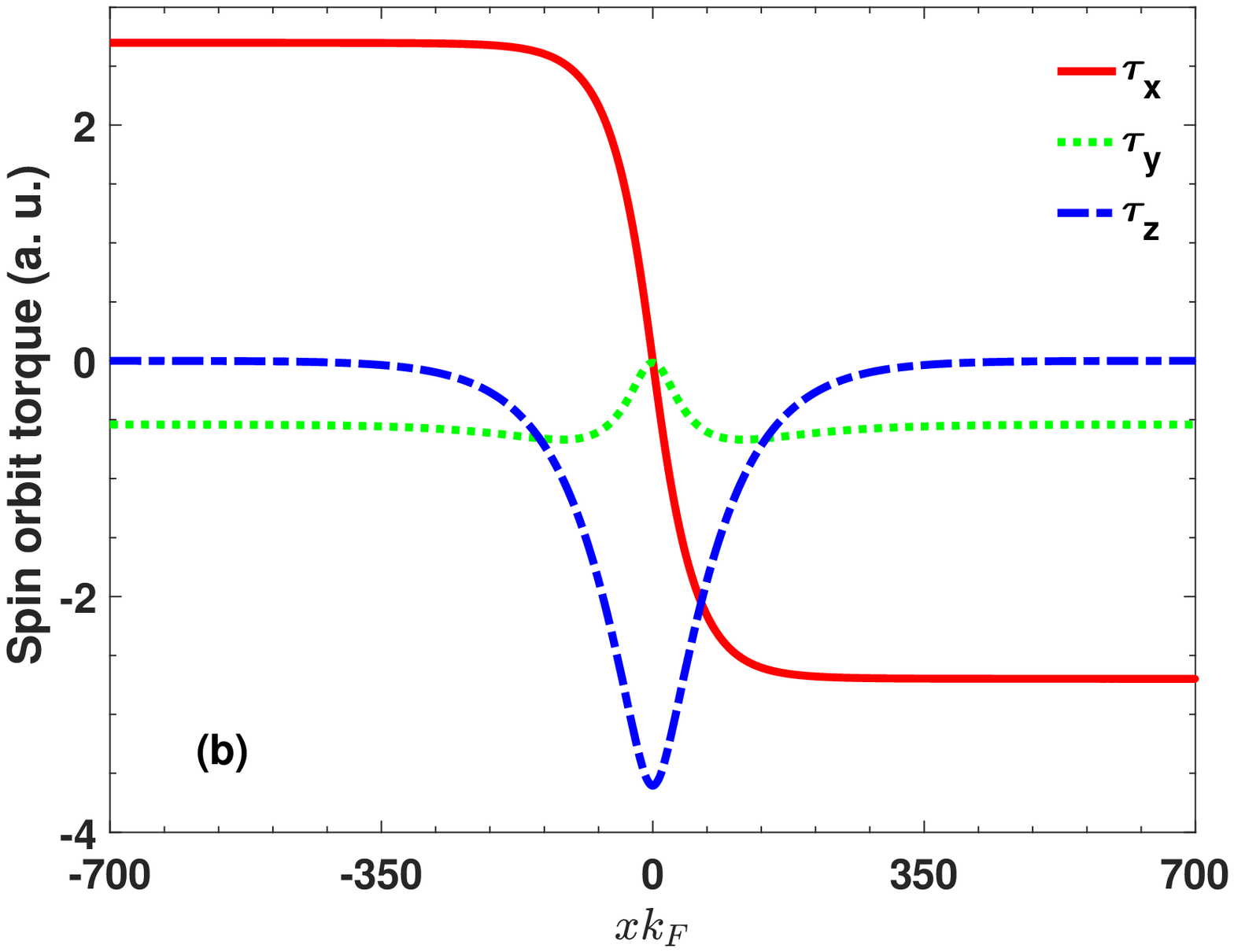}
\end{minipage}
\caption{RSOT with $q = 1$ and $\chi = 1$. The DW widths correspond to $\lambda k_F = 2$ (a) and $\lambda k_F = 70$ (b). For the small DW width $\lambda k_F = 2$ (a), both the precessional ($x$ and $z$) and the damping ($y$) components are comparable in magnitude. As the DW width increases to $\lambda k_F = 70$ (b), the damping component decreases in comparison to the precessional one. For the long DW width $\lambda k_F = 70$, although the damping component is negligibly small at the DW center, its magnitude is sizable far away from the DW center.}
\label{sot}
\end{figure}

We adopt a scattering matrix method \cite{Xia06,Zwierzycki08} to numerically solve the eigenvalue problem
\begin{equation}
H \psi = \epsilon _{\textbf{k}} \, \psi
\label{eigen}
\end{equation}
for the Pauli-Schr\"{o}dinger equation with energy $\epsilon _ {\textbf{k}}$. The idea behind this scattering matrix method is intuitively simple. In order to construct the eigenfunctions of Eq. (\ref{eigen}), we first solve it at infinity to obtain the asymptotic wave functions with specific momentum and spin. Then we evolve the obtained asymptotic wave functions towards the DW center, according to Eq. (\ref{eigen}). Generally, the evolved wave functions are not continuous at the DW center, and are thus not eigenfunctions in the whole space. This problem can be overcome by forming linear combinations of the evolved wave functions with the same energy but different momenta and spin projections along the $z$ direction, requiring that the continuity condition is satisfied at the DW center. The resultant wave functions are eigenfunctions over the whole space. Previously, the same method was successfully applied to the discussion of STT in DWs \cite{Xiao06}. In the actual numerical implementation, we can employ a particle-hole or charge-parity-time-reversal symmetry of the Hamiltonian (\ref{hamil}), $H = \sigma_x {\cal{PT}} H {\cal{TP}} \sigma_x$, to reduce the number of the wave functions to be computed. Wave functions related to each other by the particle-hole symmetry, $\psi$ and $\sigma_x {\cal P} \psi$, are conjugate pairs with opposite momenta but identical spin projections along the $z$ direction, injecting from opposite ends of the DW. It is interesting to note that a similar particle-hole symmetry was found for magnons inside DWs \cite{Wang17}. Further numerical details of the calculation are given in Ref. \onlinecite{Wang19}.

\begin{figure}
\begin{minipage}[c]{0.45\linewidth}\centering
\includegraphics[width=\linewidth]{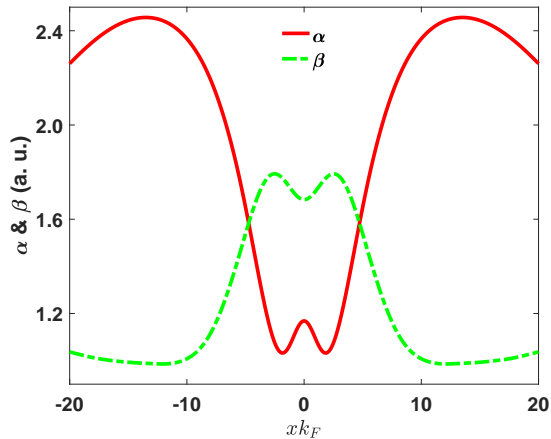}
\end{minipage}
\caption{Precessional ($\alpha$) and damping ($\beta$) RSOT coefficients with $q = 1$ and $\chi = 1$. The DW width is $\lambda k_F = 2$. The quantum confinement induced oscillation around the DW center ($x k _F$ = 0) and far away from the DW center ($x k _F$ = $\pm$ 20), which is obvious for the displayed DW width, is smoothed out as the DW width is increased to $\lambda k_F = 70$, as shown in Fig. \ref{abtopo}. Unspecified parameters are the same as those used to generate Fig. \ref{sot}.}
\label{ab}
\end{figure}

\begin{figure}
\begin{minipage}[c]{0.45\linewidth}\centering
\includegraphics[width=\linewidth]{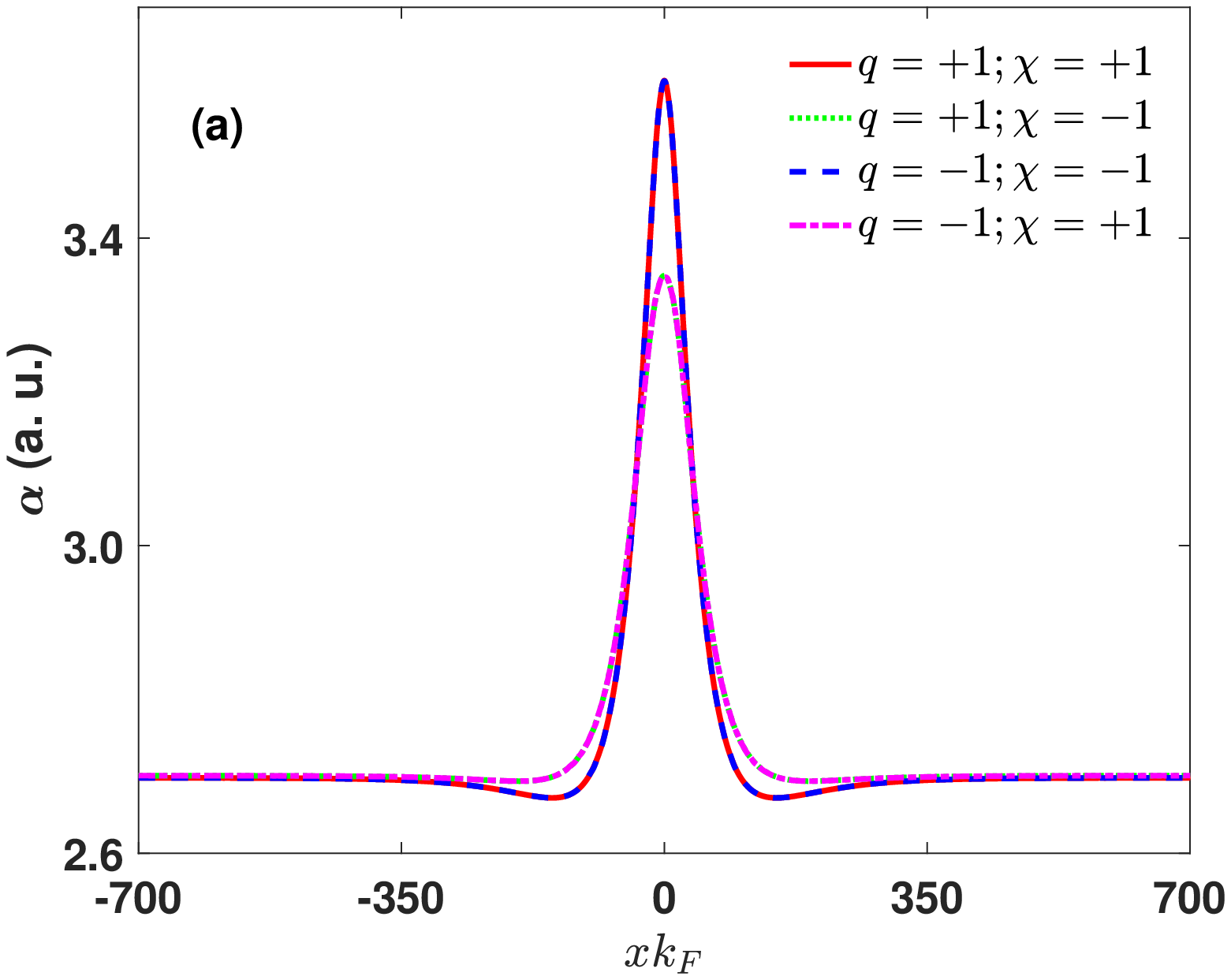}
\end{minipage}
\begin{minipage}[c]{0.45\linewidth}\centering
\includegraphics[width=\linewidth]{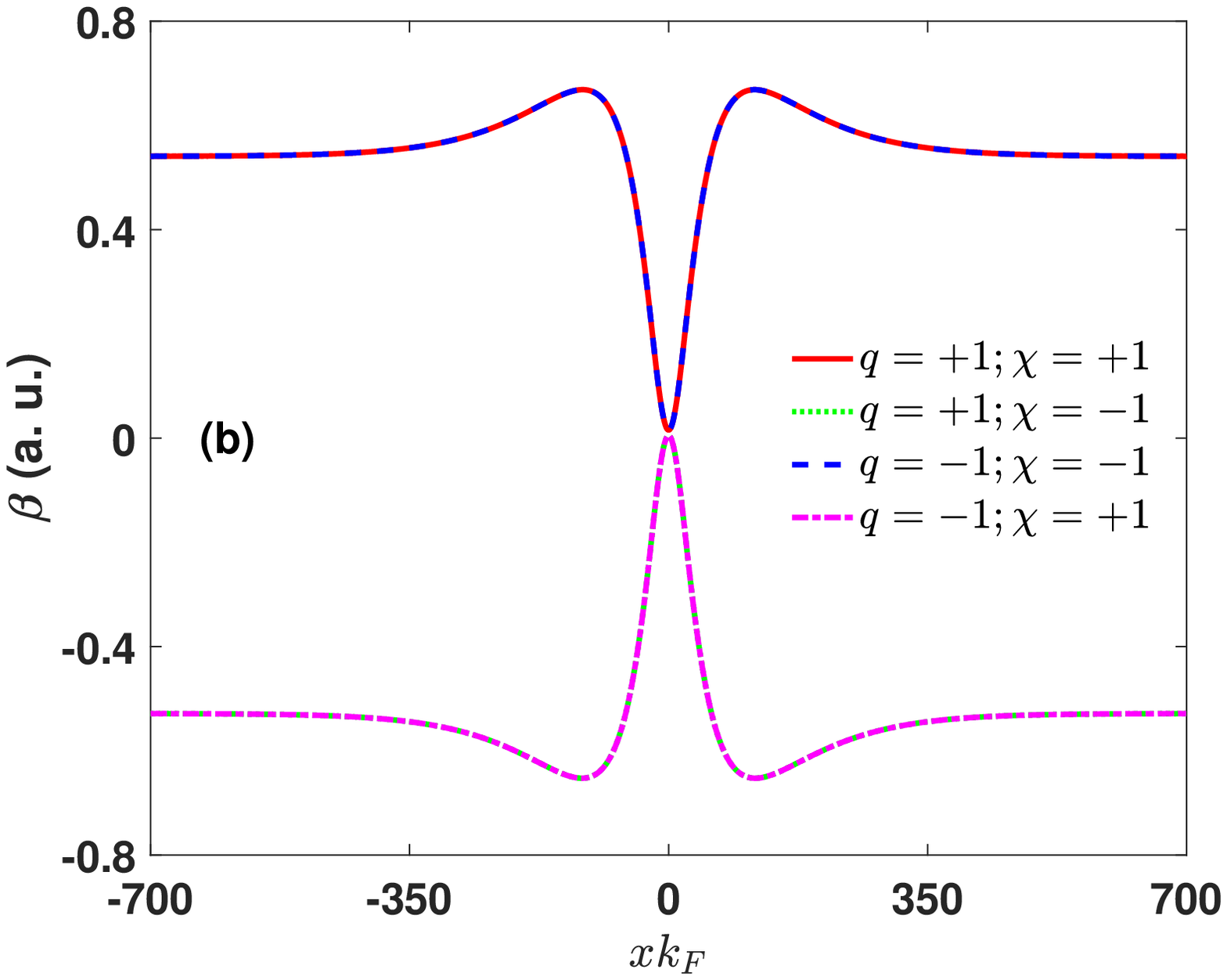}
\end{minipage}
\caption{Topological behaviour of the precessional ($\alpha$) and damping ($\beta$) RSOT coefficients for all four combinations of $q$ and $\chi$. The DW width is $\lambda k_F = 70$. Other parameters are the same as those used to generate Fig. \ref{sot}.}
\label{abtopo}
\end{figure}

With the numerical wave functions thus obtained, the RSOT can be computed using Eqs. (\ref{tau}) and (\ref{atau}). The resultant RSOT for the DW width $\lambda k_F$ = 2 and $\lambda k_F$ = 70 with $k_B/k_F = 0.4$ and $k_\alpha /k_F = 0.1$ is shown in Fig. \ref{sot}, where we have measured the DW width in terms of the inverse Fermi wave vector $k_F ^{-1}$ for the free electron system that is described by only the kinetic energy term in the Hamiltonian (\ref{hamil}). For the shorter DW width $\lambda k_F$ = 2, the RSOT has both sizable precessional and damping components. The precessional component is caused by the effective Rashba field, which has the form $\hat {m} \times \hat{y}$, while the corresponding damping component is $\hat {m} \times (\hat {m} \times \hat{y})$ \cite{Manchon}. The total RSOT is a sum of both components,
\begin{equation}
\bm {\tau} = \alpha \hat {m} \times \hat{y} + \beta \hat {m} \times (\hat {m} \times \hat{y}).
\label{tso}
\end{equation}
The corresponding decomposition coefficients $\alpha$ and $\beta$ are displayed in Fig. \ref{ab}. Due to the confinement of electrons caused by such short DWs, quantum interference of wave functions shows up as the observable spatial variation of the RSOT and decomposition coefficients far away from the DW center. This spatial variation decays out as the DW width is increased (cf. Figs. \ref{sot} (a) and (b), \ref{ab} and \ref{abtopo}).

As the DW width increases, the magnitude of the precessional component increases while the magnitude of the damping component decreases, as can be expected from a previous investigation on STT \cite{Xiao06}. However, the scaling of the non-adibaticity for the RSOT, which is defined as $\beta/\alpha$, is algebraic instead of exponential \cite{Wang19}. At the DW center, the residue damping component is negligible, but it is sizable far away from the DW center, as evident from Fig. \ref{abtopo} (b) for the longer DW width $\lambda k_F$ = 70. This finite residue damping component of the RSOT will demonstrate itself in the current driven magnetization dynamics of magnetization textures, and warrants further attention in considering its effects in spintronic devices. Furthermore, our numerical result shows that the coefficient $\beta$ depends on the topology of the underlying DW. As shown in Fig. \ref{abtopo}, for the four possible combinations of the DW charge and chirality, we have only two traces for $\beta$, reversed to each other, for the longer DW width $\lambda k_F$ = 70: the product of the DW charge and chirality, $q \chi$, determines the sign of $\beta$.

The physical origin of the damping RSOT can be determined through a perturbation analysis of the same Pauli-Schr\"{o}dinger equation (\ref{eigen}) which is used for our numerical simulation. Using the first order wave function, the damping RSOT component at $x = \pm \infty$ can be calculated. It has the form as given in Eq. (\ref{tso}) with the coefficient \cite{Wang19}
\begin{equation}
\beta \propto q \chi k_ \alpha ^2 \left( c + \frac {a} {\lambda^2} + b e^ {-\gamma \lambda} \right)
\label{beta}
\end{equation}
to the lowest order in $k_ \alpha$, where $a$, $b$, $c$ and $\gamma$ are all constants. The constant $c$ is of the order of unity, hence as the DW width increases to a very large value, $\lambda \gg \lambda_c$, the damping RSOT will approach to a constant value $c$ at $\pm \infty$. The critical length $\lambda_c = k_F/k_B^2$, which is $\lambda_c k_F = 6.25$ using our parameters, determines the DW width where transition from non-adiabatic to adiabatic behaviour occurs for STT in DWs without spin orbit interaction \cite{Xiao06}.

The appearance of the factor $q \chi$ in the expression of $\beta$ indicates that the damping RSOT is a topological quantity. The factor $k _\alpha ^2$ signifies that the damping RSOT is a higher order effect, as $k _\alpha$ is proportional to $\alpha_R$. In the perturbation calculation, the adiabatic or zeroth order wave functions give rise to only the precessional RSOT. Due to this origin from the zeroth order wave functions, the adiabatic coefficient $\alpha$ is almost independent of the topological features of the underlying DW, whether in the adiabatic limit or not: For $\alpha$, the dominant contribution does not sense the topology of the DW, and the topological contribution only enters as a higher order correction (cf. Fig. \ref{abtopo} (a)). Inclusion of the first order wave functions brings about the damping RSOT. The first order wave functions at infinity are determined by the scattering of the incident, zeroth order waves under the influence of the perturbation potential. To the first order of $k_ \alpha$, the explicit form of the perturbation potential in momentum space $V(k_f, k_i)$ for incident and scattered momenta $k_i$ and $k_f$ is give by
\begin{eqnarray}
V(k_f, k_i) &=& \frac {p \, \mbox {csch} p} {4 \pi \lambda} - \chi \frac {k_ \alpha} {4} \frac {k_y} {k_B^2} \frac {\pi^2 + 4 p^2} {2 \pi^2 \lambda} \mbox {sech} p + q \chi \frac {k_ \alpha} {4} \mbox {sech} p \nonumber\\
&-& q \chi \frac {k_s} {4} \left(\mbox {sech} p - 2 \chi \frac {k_ \alpha k_y} {\pi k_B^2} p \, \mbox {csch} p \right) \sigma_y + \chi \frac {\lambda k_ \alpha} {2} k_y \sigma_z \mbox {sech} p,
\label{vk}
\end{eqnarray}
with $p = Q \lambda \pi/2$ and $k_s = k_f + k_i$. $Q = k_f - k_i$ is the momentum transfer. In comparison to the original potential in (\ref{hamil}), the potential (\ref{vk}) corresponds to a magnetic field with only $y$ and $z$ components and a scalar electric potential, while the Rashba interaction is absorbed into the magnetic field and electric potential. When the momentum transfer is zero, the scaling of $V(k_i, k_i)$ with respect to the DW width $\lambda$ is algebraic. For finite momentum transfer $Q = k_f - k_i$, $V(k_f, k_i)$ brings about the exponential decay of the physical quantities on the DW width through the hyperbolic secant and cosecant functions \cite{Dugaev02}.

Not all of the topological terms in potential (\ref{vk}) contribute to the expression for the damping RSOT. In the case of zero momentum transfer, $Q = 0$, the $y$ component of the effective magnetic field in (\ref{vk}), which is the coefficient of $\sigma_y$, does not contribute at all; while the $z$ effective field, which is the coefficient of $\sigma_z$, and the scalar potential contribute partly: The product of the first and second terms in the scalar potential gives rise to the term proportional to $\lambda ^ {-2}$ in (\ref{beta}), while the product of the $z$ component of the effective magnetic field with the first term of the scalar potential contributes the constant term in $\beta$. Both those two contributions are proportional to the chirality $\chi$. The dependence of $q$ in the final expression for $\beta$ is derived from its dependence on the $z$ component of the magnetization, $m_z$. Hence the topological feature of $\beta$ is characterized by the relation $\beta \propto \chi m_z$, far away from the DW center. This behaviour is similar to that of Bloch wave functions in periodic potentials, as demonstrated by the Zak phase \cite{Zak}. The $m_z$ is a dynamical contribution, and $\chi$ is a manifestation of the existence of a topological phase with the value of $0$ or $\pi$. The topological dependence of $\beta$ obtained using the potential (\ref{vk}), Eq. (\ref{beta}), is actually borne out by the numerical results, as shown in Fig. \ref{abtopo}.

The topological nature of $\beta$ explains mathematically why the damping RSOT remains finite even when the DW width is large, $\lambda \gg \lambda_c$. Due to the different topologies of the DW and a uniformly magnetized state, a continuous transition between the two states is prohibited. Thus $\beta$ cannot be reduced to zero, which is the value for $\beta$ in a uniform state. Physically, the topological protection of the damping RSOT can be traced back to the nonlocal character of quantum particles, which means that the wave functions are not determined locally by the potential. In particular, the damping RSOT at $x = \pm \infty$ is determined by $V(k_i, k_i)$ in the adiabatic limit ($\lambda \gg \lambda_c$), which is an integration of the perturbation potential over the whole space and gives rise to the topological characteristics of the damping RSOT. Therefore, the damping RSOT at $x = \pm \infty$ is finite due to the pure existence of the DW, even though the magnetization variation there is infinitesimal, approaching to the value for a uniform magnetization distribution.

To see how the newly identified damping RSOT influences the current driven DW motion (CDWM), we consider the expression for the normalized DW velocity $v$
\begin{equation}
n \alpha_G v = q h_z \cos \theta _h - n \xi u + q \chi \beta \Delta \theta _0,
\label{dw_v}
\end{equation}
obtained using a simple 1D model description of CDWM \cite{Wang16}. A detailed derivation of the velocity (\ref{dw_v}) is given in the Supplemental Material [\onlinecite{supp}], and references [\onlinecite{Hickey}, \onlinecite{Slastikov}] therein. $\alpha_G$ is the Gilbert damping constant, and $n$, $\theta_h$ and $\Delta \theta _0$ are constants related to the equilibrium DW configuration. $\xi$ is the non-adiabaticity and $u$ is an equivalent speed for the STT. $h_z$ is a perpendicularly applied magnetic field, normalized to the anisotropy field. In obtaining Eq. (\ref{dw_v}), we have assumed that the current density is small, and the RSOT only causes infinitesimal deviation from the equilibrium DW configuration. Even with this rather simple assumption, Eq. (\ref{dw_v}) shows that the CDWM can exhibit very complicated behaviour: For short DWs, $\beta$ is not completely determined by the product $q \chi$, then the RSOT contribution to the DW velocity has both $q \chi$ dependent and independent components. In the adiabatic limit, the $q \chi$ dependent component of $\beta$ fades out, and the RSOT contribution to the DW velocity is $q \chi$ independent, resembling the behaviour of the STT contribution.

In systems with a sizable Rashba interaction, the sign of the product $q \chi$ for a stable N\'{e}el wall is determined by the sign of the Dzyaloshinskii-Moriya interaction constant $D$, as $q \chi D <0$ gives a lower energy. Additional control over the DW chirality can be realized by applying an in-plane magnetic field $h_x$ \cite{Haazen}, with the DW charge fixed. With this freedom in manipulating the DW chirality, the velocity of CDWM can be tuned by the application of an in-plane magnetic field, for DW width in the non-adiabatic limit. Further complication can arise from the chirality dependence of the Gilbert damping constant and gyromagnetic factor \cite{Jue,Akosa,Freimuth17}, as well as the STT non-adiabaticity \cite{Wang19STT}. Before turning to our conclusion, it is appropriate to mention that our above discussion is based on a simple 1D treatment of the CDWM, which is a very rough approximation based on the assumption that the ground state of the DW is a N\'{e}el configuration. The applicability of this assumption is dubious in the presence of an electric current, since the current induced effective Rashba field tends to stabilize a Bloch wall. Our discussion is only to illustrate the complication of the CDWM in the presence of the RSOT. Further detailed investigation is needed for a thorough understanding of the CDWM in systems with sizable Rashba spin orbit interaction.

In conclusion, we have studied the RSOT in magnetic DWs, which is derived from the broken inversion symmetry at ferromagnet/heavy metal interfaces. By numerically solving the Pauli-Schr\"{o}dinger equation for 2D electrons moving inside a N\'{e}el DW, a topological damping RSOT component is identified. Even in the adiabatic limit, the magnitude of the topological damping component is sizable, in stark contrast to the negligible non-adiabatic STT in the same limit. This finite damping RSOT is a manifestation of the nontrivial topology of the underlying DW. The identification of a topological damping RSOT component in magnetization textures will promote the application of RSOT in spintronic devices and facilitate a thorough understanding of the experimental data in current driven motion of magnetic DWs in ferromagnet/heavy metal bilayer systems.

\section*{Acknowledgements}
We would like to express our gratitude to Prof. Jiang Xiao for his valuable comments and discussions, especially for bringing us to the topic of RSOT in magnetic DWs and sharing his code on STT simulation. Y. Z. acknowledges the support by the President's Fund of CUHKSZ, Longgang Key Laboratory of Applied Spintronics, National Natural Science Foundation of China (Grant No. 11974298, 61961136006), and Shenzhen Fundamental Research Fund (Grant No. JCYJ20170410171958839).

\end{document}